# RXTE Observation of the Nonthermal Emission from the Early Stage Merger in A1750

Mark J. Henriksen
Physics Department, University of Maryland, Baltimore County, Baltimore, MD 21250, USA; henrikse@umbc.edu

## Abstract

We make the first observation-based calculation of the energy that goes into cosmic ray protons versus cosmic ray electrons in shock acceleration during structure formation. We find a ratio of energy in cosmic ray protons to energy in cosmic ray electrons of 0.86. This value, calculated from the nonthermal X-ray component reported here from RTXE and the Fermi LAT upper limit for gamma-ray emission, is significantly lower than theoretical estimates that place most of the nonthermal energy in protons. Our estimate is based on the detection of nonthermal X-ray emission using the 3–20 keV RXTE spectrum, which shows residual emission not well modeled by a single thermal component. The statistical significance of adding a nonthermal, power-law component is 96%. The significance of adding a second thermal component is 90%. The addition of a component consisting of full cosmic X-ray background fluctuation to an isothermal model is significant with 92% confidence. The cumulative probability for the two-thermal-component model is 81% and 90% for the thermal plus power law. Thus the model with nonthermal emission is the preferred description of the data. Evidence of shock heating between the clusters in the spectro-imaging data of XMM, Chandra, and Suzaku indicates that a cosmic ray component should also be present and supports a nonthermal interpretation for the additional component. The bolometric nonthermal X-ray luminosity is $1.6 \times 10^{44}$ ergs s$^{-1}$, 36% of the total X-ray emission in the 0.1–100 keV band.

*Unified Astronomy Thesaurus concepts:* Non-thermal radiation sources (1119); Galaxy clusters (584)

## 1. Introduction

Galaxy clusters form from the merging of smaller subclusters. Simulations show that multiple small subclusters merge to form two nearly equal subclusters, which then merge to form the final massive cluster, which may then experience continued small accretions (Wu et al. 2013). Massive merging occurs frequently in clusters observed several billion years ago when these young structures are experiencing their strongest formation event. When merging occurs between subclusters of nearly equal mass, or repeated accretion of smaller masses, a large quantity of energy ($10^{63}$–$10^{64}$ ergs) is released in the intracluster medium (hereafter ICM). This energy budget heats the ICM gas nonuniformly through shocks produced by the mergers (Pavlidou & Fields 2006). Observational evidence of temperature structure in nearby clusters has been accumulated using ROSAT, ASCA, Chandra, Suzaku, and XMM observations. An apparent trend is that nearly every cluster that shows a nonuniform temperature distribution is also found to be undergoing a merger (Vikhlinin et al. 2005). A1750 is a powerful and spectacular early stage merging cluster (Forman et al. 1981; Donnelly et al. 2001) with visible compression and heating at the subcluster interface in the XMM temperature and surface brightness maps (Belsole et al. 2004), Chandra (Molnar et al. 2013), and Suzaku (Bulbul et al. 2013). However, temperature maps are fundamentally only indirect probes of such merging phenomena in clusters and may be ambiguous depending on the location in the cluster. This is because heating of the ICM could be due, in principle, to other mechanisms (e.g., active galactic nucleus (AGN) outflow; Nulsen et al. 2007, galactic winds; Mitall et al. 2009). A direct way to identify the existence of accretion shocks is through nonthermal emission. The interaction of shocks with thermal gas is affected by electromagnetic fields in the ICM. Particles in the thermal distribution are accelerated to high energy via diffusive shock acceleration (Blandford & Eichler 1987) forming a nonthermal component. The result is that a fraction of the kinetic energy of the shock is converted into cosmic rays (Kang et al. 2002). However, GeV cosmic ray electrons have a short cooling time, $\sim 10^8$ yr, due to inverse Compton emission and/or synchrotron emission in a micro-Gauss magnetic field (Enßlin et al. 2011). Thus, they require reacceleration. Reacceleration of electrons by compressible turbulence induced by cluster mergers has been suggested to explain the reported radio halos and relics in clusters (Brunetti & Lazarian 2007). A strong, high Mach number shock and weak accretion shocks are predicted to produce cosmic rays with conversion efficiencies ranging from a few percent up to $\sim 50\%$ (Ryu et al. 2003). But upper limits from gamma-ray observations of galaxy clusters with Fermi that exclude hadronic injection efficiencies above 21% (Ackermann et al. 2014) may indicate that the simulations are overestimating the efficiency of cosmic ray acceleration from mergers, with cosmic ray to shock energy ratios of less than 1%, being more realistic.

While primary cosmic ray electrons, those accelerated at the shock front, produce synchrotron and inverse Compton emission near the shock, the cosmic ray protons diffuse outward, colliding with ions in the ICM with an average time between collisions of 10 Gyr in the low density regions and shorter in the cluster cores where the density is higher. AGNs fail to provide sufficient heating to balance cooling in the cluster core (Mitall et al. 2009) and the interaction of proton cosmic rays with core gas may provide the missing heating component (Berrington & Dermer 2003). This is important because substructure in clusters is anticorrelated with the presence of a cool core (Bohringer et al. 2010).

The cosmic ray protons produce gamma-rays through $pp \rightarrow \pi^0 \rightarrow \gamma + \gamma$, and secondary electrons through $pp \rightarrow \pi^{\pm} \rightarrow \mu^{\pm} \rightarrow e^{\pm} + \bar{\nu}_{\mu}(\nu_{\mu}) + \nu_{e}(\bar{\nu}_{e})$. The secondary electrons, in turn, produce large diffuse radio halos or mini halos via the synchrotron process in a $\sim\mu G$ intracluster magnetic field (Pfrommer et al. 2008) and nonthermal X-rays through inverse Compton scattering with the cosmic microwave background photons. These analytical studies have been confirmed by numerical simulations that have clearly established a physical connection between the radio halo flux, nonthermal X-ray emission, and merger shock (Miniati et al. 2001; Ha et al. 2018).

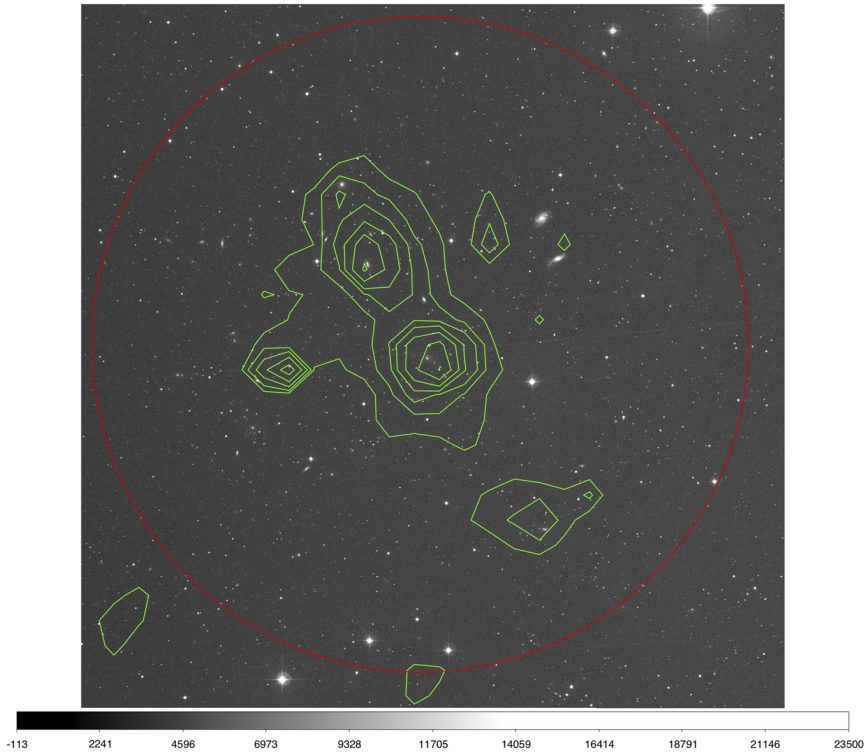

**Figure 1.** This shows what part of the X-ray cluster is in the RXTE PCA beam.

The energetics of cosmic ray acceleration via merger shocks is plausible for producing several nonthermal emission components, each with substantial luminosity: (1) radio halos, (2) $\gamma$-rays through hadronic processes, and (3) X-ray and $\gamma$-ray emission via inverse Compton from primary and secondary cosmic ray electrons. Radio halos appear ubiquitous in cluster mergers with a frequency of 90% (Cuciti et al. 2021). The fraction of energy that goes into cosmic rays can be estimated through both observations and simulations. The kinetic energy from an accretion event in a rich cluster is $\sim \rho V_{\mathrm{flow}}^3 R_{\mathrm{cl}}^2$ (Miniati et al. 2001). In the case of A1750, the flow velocity in the initial shock is approximately equal to the merger velocity 1460 km s$^{-1}$, derived from hydrodynamical simulations of the merger to be consistent with the X-ray observations and the projected velocity dispersion, 900 km s$^{-1}$ (Molnar et al. 2013). This gives $7.2 \times 10^{45}$ ergs s$^{-1}$ of kinetic energy through the shock. The adiabatic sound speed is $\sim 618$ km s$^{-1}$ for a temperature of $\sim 2.4$ keV at the virial radius (De Grandi & Molendi 2002; Vikhlinin et al. 2005; Leccardi & Molendi 2008; Bulbul et al. 2013). The fractional conversion to cosmic rays for a weak shock, Mach $\sim$ 2–3, is $\sim$5% (Kang et al. 2002). Multiplying by the kinetic energy from the merger predicts $\sim 3.6 \times 10^{44}$ ergs s$^{-1}$ in cosmic rays. Modifications include preheating of the cluster atmosphere to 1 KeV (Borgani et al. 2008), which may provide a lower limit to the temperature in the outer atmosphere. In that case, accretion shocks could be considerably stronger.

The relative fraction of energy going into cosmic ray protons versus electrons, which we will call K, is unknown and will affect the predicted gamma-ray and X-ray luminosities, which are known. Simulations have utilized $\sim$10 on the grounds that protons resonate at all energies with Alfvén waves and that the Larmor radius of protons in the thermal distribution is larger than for electrons, thus allowing them to more efficiently scatter across the shock (Baring 1999; Berrington & Dermer 2003). The realization of this prediction will lead to a strong $\gamma$-ray signature from protons in merging clusters. The value of K will vary for any particular cluster depending upon the energetics of the merger. The accuracy of the conversion of shock energy into cosmic rays used in simulations can be measured through comparison of the nonthermal X-ray luminosity (as we report here, which is due to cosmic ray electrons) and $\gamma$-ray detection, which is due to protons (Pinzke & Pfrommer 2010). As there are only upper limits on the gamma-ray emission from galaxy clusters (Ackermann et al. 2014; Griffin et al. 2014), the value on K will also be an upper limit. In the discussion section we calculate this parameter using the Fermi LAT upper limit for stacked clusters and the nonthermal X-ray detection reported here for A1750.

## 2. Observations and Data Analysis

The RXTE Proportional Counter Array (PCA) observed the A1750 cluster for 214,000 s. The PCA on board RXTE is a collimator with a 1° FWHM field of view (FOV). Figure 1 shows the PCA FOV as a red circle and the X-ray contours from the wide FOV PSPC detector overlaid on the optical image. This image shows that the majority of the X-ray emission is in the RXTE FOV. Data from the top layer of PCA array 2 was used to extract the highest signal-to-noise spectrum. Filtering out bright Earth, slews, high electron contamination, and times near the South Atlantic anomaly was done to select good time intervals. Finally, a count rate filter was applied to the light curve to eliminate spikes and obtain a flat light curve for the extracted spectrum. The resulting PCA spectrum's energy band is 3–30 keV with a background subtracted count rate of $1.43 +/- 0.013$ counts s$^{-1}$ (cps).

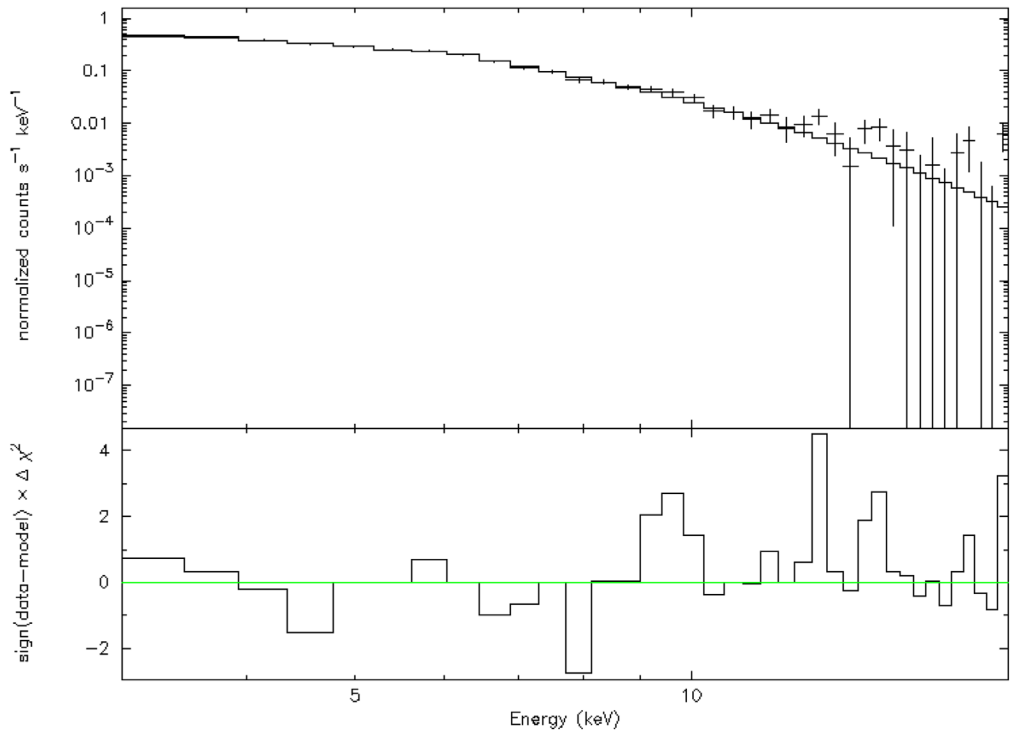

**Figure 2.** Systematic residuals in the PCA at $E > 5$ keV for the thermal model.

There are several PCA specific calibration issues that are potentially relevant. The PCA dead time correction is important in matching the absolute flux of the Crab. The estimated dead time is $10^{-5}$ s per event. For a high count rate object, such as the Crab, 13,000 cps, this is 0.13 dead seconds per second and can lead to an underestimate of the Crab flux by 6% if not accounted for. However, A1750, with a gross PCA count rate of 15.2 cps, has a total dead time of 32 s over the 214 ks observation, which is negligible for the absolute flux calibration. Additional tweaking of the effective areas to match the Crab flux has been included in the version of xpcaarf in HEASOFT 6.9 tools, which were used in this analysis. Accurate modeling of Xenon L-escape features, a better channel to the energy conversion table, and the requirement of a better fit to the Crab canonical spectrum for the PCA has reduced the systematic energy calibration error to 0.5% (Shaposhnikov et al. 2009).

Models are fit to the spectrum using the spectral analysis package, Xspec (Arnaud 1996). The primary models consist of the following components: single thermal, two thermal, and thermal plus a power law. The absorbing column of neutral hydrogen is a common parameter to all models. This parameter is fixed at $2.71 \times 10^{20}$ cm$^{-2}$, the cluster average obtained with the HEASARC n$_H$ tool. The thermal components are the Raymond model described in the Xspec documentation, which have metal abundance and temperature as free parameters. The power-law model has the X-ray spectral index, which is fixed at the typical value of 2.2 based on cluster radio halo spectra (Cuciti et al. 2018). For the two-thermal-component fit, the high temperature component is fixed at the shocked bridge gas temperature, 5.12 keV, for the subcluster merger (Belsole et al. 2004), and the abundance is required to be the same for both thermal components. These restrictions are necessary for the model fitting to converge. All components have a normalization, which is proportional to the product of the electron and hydrogen ion density, integrated over volume for thermal components, and for the power law, the normalization is proportional to the cosmic ray electron density. Since a model consisting of two temperatures is a better fit than a single temperature, we refit the spectrum with a model consisting of three components: the power law added to the two-thermal-component model. This adds an additional free parameter: The normalization of the nonthermal component. Table 3 has the results of fitting this three-component model to the spectrum. The minimum $\chi^2$ improves with the addition of the nonthermal component supporting the presence of nonthermal emission. However, there is no improvement over the model consisting of a single thermal and a nonthermal component because in this three-component fit, the second thermal component is essentially marginalized. We conclude that the spectrum is not sufficient to constrain additional components beyond the thermal plus nonthermal model. There are residual spatial fluctuations with a maximum of 7%–8% in the hard cosmic X-ray background (CXB) due to unresolved point sources. The CXB contribution was modeled as a power law with index −1.29, cutoff at 41.13 keV, and variable amplitude with a maximum of 8% of the mean CXB flux at 20 keV (normalization of $+/-1.84 \times 10^{-4}$).

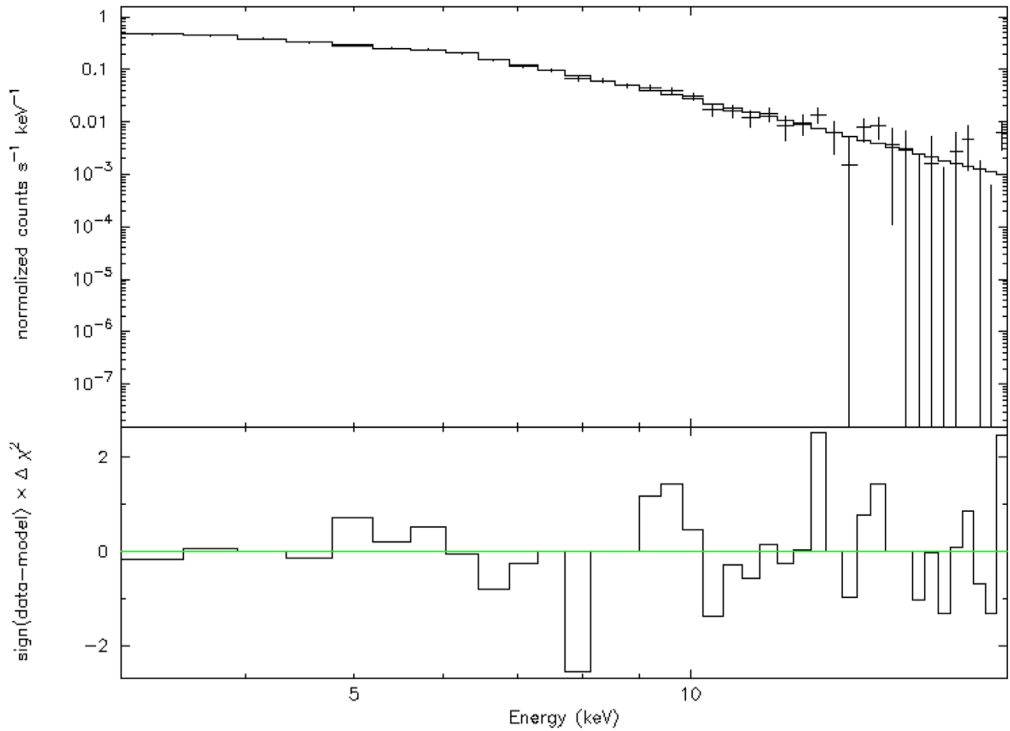

**Figure 3.** Addition of a power law significantly diminishes the amplitude of the residual emission above 5 keV.

**Table 1**
A1750 Fit Parameters

| Model | kT | Abundance | Spectral Index | kT | $\chi^2$/dof |
|---|---|---|---|---|---|
| Thermal | 3.66–3.96 | 0.13–0.21 | ⋯ | ⋯ | 33.7/36 |
| Thermal + Power Law | 2.86–3.57 | 0.19–0.36 | 2.2 | ⋯ | 24.9/35 |
| Thermal + Thermal | 1.83–3.23 | 0.16–0.26 | ⋯ | 5.12 | 27.6/35 |
| Thermal + Full CXB | 2.87–3.11 | 0.21–0.31 | ⋯ | ⋯ | 26.6/36 |

**Table 2**
A1750 Fit and Calculated Parameters

| Model | Thermal Norm First | Flux (3–20 keV) $\times 10^{-12}$ | $L_x$ (3–20 keV) (0.1–100 keV) $\times 10^{44}$ | Thermal/NT Norm Second | Flux (3–20 keV) $\times 10^{-12}$ | $L_x$ (3–20 keV) (0.1–100 keV) $\times 10^{44}$ |
|---|---|---|---|---|---|---|
| Thermal | 0.016–0.017 | 6.7 | 1.0 (3.2) | ⋯ | ⋯ | ⋯ |
| Thermal + Power Law | 0.014–0.016 | 5.0 | 0.7(2.8) | 0.000468–0.00150 | 2.2 | 0.3 (1.6) |
| Thermal + Thermal | 0.011–0.013 | 2.1 | 0.3 (1.7) | 0.004–0.0088 | 4.8 | 0.7 (1.9) |
| Thermal + Full CXB | 0.018–0.0190 | 7.4 | 1.0 (4.4) | ⋯ | ⋯ | ⋯ |

## 3. Results: Nonthermal Emission

The first model fit, a single thermal component, is a good fit with a reduced chi-square of 0.91 (See Figure 2). The F-test is used to compare more complex models. Adding a power-law component with spectral index fixed at 2.2 is a significant improvement of the fit with 96% confidence (See Figure 3). Similarly, instead adding a second thermal so that the model consists of two thermal components, is a significant improvement over a single thermal with 90% confidence. There are fluctuations in the hard X-ray background that may be in the PCA FOV. The fluctuations may be as large as +/−8% of the hard X-ray background (HXB). If a full, positive, X-ray background fluctuation is added to a single thermal component, the worse case for detecting nonthermal emission, is an improvement of 92%. While there is a slight preference for a nonthermal component identified with the cluster, a second, harder thermal component or even a strong HXB fluctuation is of a similar likelihood based solely on data analysis (See Table 1).

For $H_0 = 67.74$, $\Omega_m = 0.309$, $\Omega_\Lambda = 0.691$ (Planck Collaboration et al. 2016), and redshift, $z = 0.0852$ (Struble & Rood 1999), the luminosity distance is 406 Mpc (Wright 2006). We calculated the luminosity of all components in the 3–20 keV band of the RXTE PCA using the luminosity distance (See Table 2). We compare the RXTE luminosities to XMM, 0.3–10 keV, (Belsole et al. 2004) after converting both to the bolometric X-ray band, 0.1–100 keV. The major XMM components, computed using best-fit parameters, give a total luminosity, the sum of the three XMM components, of $4.4 \times 10^{44}$ ergs s$^{-1}$. The XMM luminosity is in perfect agreement with the PCA's thermal plus power law and thermal plus full CXB models, 18% higher than two thermals, and 27% higher than the single thermal. This indicates that the models of a single thermal with full CXB and thermal plus power law each contain all of the cluster emission observed with XMM imaging. Table 3 has the results of fitting the three-component model to the spectrum. The minimum $\chi^2$ improves with the addition of the nonthermal component supporting the presence of nonthermal emission. However, there is no improvement over the model consisting of a single thermal and a nonthermal component because in this three-component fit, the second thermal component is essentially marginalized. We conclude that the spectrum is not sufficient to constrain additional components beyond the thermal plus nonthermal model.

Table 3
A1750: Three-component-fit Parameters

| Model | kT | Abundance | Low $L_x$ (0.1–100 keV) 3–20 keV × $10^{43}$ | NT $L_x$ (0.1–100 keV) 3–20 keV × $10^{43}$ | High Thermal $L_x$ (0.1–100 keV) 3–20 keV × $10^{43}$ | $\chi^2$/dof |
|---|---|---|---|---|---|---|
| 2 Thermal + Power Law | 2.22–3.54 | 0.19–0.36 | 2.8 (11.0) | 7.0 (28.0) | 0.0024 (0.012) | 24.8/34 |

## 4. Limits on Contaminating Emission from Radio Sources in the X-Ray

There are a number of radio sources in the RXTE PCA field of view (1° FWHM); however, there are no extended sources that could be called radio halos or relics, only point sources. The NVSS survey shows that the radio flux in that region is almost entirely identified with NGC 5183 (13.4 mJ) and NGC 5184 (29.7 mJ). Using the XAMIN tool at the HEASARC, NGC 5183 has an XMM source at 13:30:5.83, −01:43:17.17 of 1.4 +/−13 cps, or a flux of $6.0 \times 10^{-14}$ ergs cm$^{-2}$ s$^{-1}$ in the (0.7–10 keV). NGC 5184 has a PSPC source at 13:30:01, −01:39:38 with 0.003 +/−0.0007 cps (0.5–2.4 keV). Using the HEASARC Pimms tool to convert both to bolometric luminosities gives $1.3 \times 10^{42}$ and $2.4 \times 10^{41}$ ergs s$^{-1}$ for 5183 and 5184, respectively. Together, they produce around 1% of the nonthermal emission, at most.

## 5. Discussion

This is a unique result in that modeling the second spectral component in the X-ray shows a preference for a nonthermal component over a thermal component. We convert the upper limit for the gamma-ray photon flux, $2.3 \times 10^{-11}$ ph cm$^{-2}$ s$^{-1}$ in the 0.8–100 GeV range (Griffin et al. 2014) using an average energy of 50 GeV/photon to an energy flux of $1.84 \times 10^{-12}$ ph cm$^{-2}$ s$^{-1}$. Using the gamma-ray upper limit and our X-ray detection for A1750, the upper limit for the ratio of the gamma-ray to X-ray flux is 0.86, from which we infer an upper limit for the energy in cosmic ray protons to cosmic ray electrons of 0.86.

Baring (1999) advanced general arguments that the overall efficiency of proton acceleration should be higher than electrons through the merger history of a cluster. For a 1% efficiency of shock energy injected into cosmic rays, the Coma cluster was predicted to be detected in GeV and TeV. Later modelers came up with similar predictions. Yet, even the stacking of tens of clusters leads to a nondetection in the GeV gamma-ray and nondetection in the TeV.

Our upper limit of ∼1 for the energy in cosmic ray protons to electron ratio, compared to assumed values of 10, indicates an overly optimistic view of the physical conditions and merger history that would favor cosmic ray proton acceleration in theoretical modeling. In addition, the strength of the shocks may be overestimated for subclusters, such as A1750, and they may merge with velocity substantially below freefall, reducing gamma-ray emission.

NuStar observations of the central region of the A523 cluster radio halo finds a 3$\sigma$ upper limit on the 20–80 keV nonthermal flux of $3.17 \times 10^{-13}$ erg cm$^{-2}$ s$^{-1}$ (Cova et al. 2019). We use the HEASARC Pimms tool to project the PCA flux into the NuStar 20–80 keV band for comparison. The A1750 flux is $1.3 \times 10^{-12}$ or 3–6 times higher. The difference might be the presence of a radio halo in A523. The central magnetic field could be enhanced by galactic processes in A523, perhaps by ram pressure stripping of infalling galaxies (Muller et al. 2021). Radio polarization studies show that the magnetic field decreases with radius in clusters (Kim et al. 1991; Govoni et al. 2006). A lower magnetic field at the subcluster interface in A1750, which is at a large distance from the cluster center, is then expected. However, the cosmic ray electron density of the early stage merger of A1750 could be considerably higher compared to the central region of A523, which may have substantial synchrotron losses since the magnetic field is much stronger.

A1750 is also somewhat unique in that there is no radio halo though radio halos are associated with cluster mergers (Cuciti et al. 2015) and are even found in weak mergers. For example, Abell 3404, has a faint, ultra-steep-spectrum radio halo (Duchesne et al. 2021). Buote (2001) find that energetically, radio halos are correlated with massive core-disrupted clusters. A1750 is in a premerger stage and is not core-disrupted; however, there is shocked gas at the interface between the subclusters and, as reported here, a population of cosmic rays. With the relative lack of pollution of the ICM, with galactic sources of magnetic fields via stripping in the low density outskirts, the field is likely closer to a primordial value. We suggest that the lack of a halo indicates that the shocked region's magnetic field is too low to produce a visible halo. It is also likely that the lack of a strong field in a premerger cluster is indicative of the merger itself and may amplify the primordial field up to a micro-Gauss; the typical field of the postmerger cluster field. The turbulent reacceleration of secondary cosmic ray electrons produced by hadronic processes can explain the existence of diffuse radio sources (Nishiwaki et al. 2021). A key element of this theory is the long-lived cosmic ray protons that diffuse throughout the cluster. The lack of a radio halo in A1750 may be due to the fact that the merger is just beginning and this process has not had time to play out. Thus, the lack of a radio halo is consistent with nonthermal emission due to primary electrons reaccelerated near the shock front.

## ORCID iDs

Mark J. Henriksen  https://orcid.org/0000-0003-0530-8736